\documentclass[aps,twocolumn,
groupedaddress,longbibliography,prb]{revtex4-2}
\usepackage{graphicx} 
\usepackage[colorlinks=true,citecolor=blue,linkcolor=blue]{hyperref}
\usepackage{dcolumn}
\usepackage{bm}
\usepackage{color}
\usepackage{tabularx}
\usepackage{comment}
\usepackage{float}
\usepackage{amsmath}
\usepackage{mathtools}
\usepackage{physics}
\usepackage{amssymb}
\usepackage{soul}
\usepackage{tcolorbox}
\usepackage{makecell}

\begin{document}

\title{General Conditions for Axis Dependent Conduction Polarity}
\author{Poulomi Chakraborty, Brian Skinner, Penghao Zhu\\
\textit{Department of Physics, The Ohio State University, Columbus, Ohio 43210, USA}}

\begin{abstract}
    Axis-Dependent Conduction Polarity (ADCP) refers to the phenomenon in which electrical transport within a single material is p-type along one crystallographic direction and n-type along the perpendicular direction. This behavior enables a variety of thermoelectric applications that do not require a heterojunction between two different materials. In this work, we investigate ADCP theoretically and derive a set of generic and quantitative criteria for identifying and predicting materials that exhibit ADCP. Specifically, by analyzing the thermopower for generic metals, semimetals, and semiconductors, we obtain transparent inequalities that are both necessary and sufficient for the emergence of ADCP. Moreover, we review known ADCP materials and verify that their band-structure characteristics and relaxation parameters are consistent with the inequalities derived here.
\end{abstract}

\maketitle

\section{Introduction}

\begin{figure}
    \centering
    \includegraphics[width=1 \columnwidth]{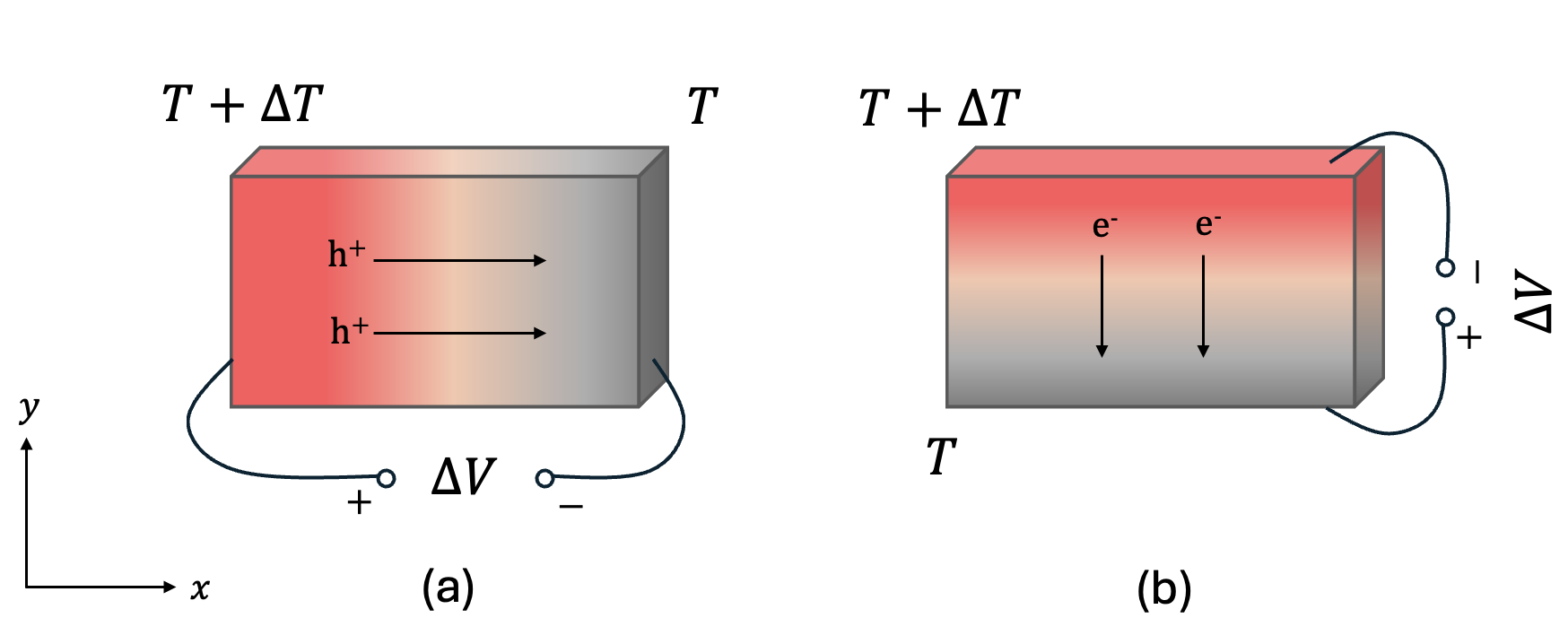}
    \caption{ Thermopower is generated when carriers (electrons/holes) move from the hot end to the cold end of the material (temperature difference $\Delta T$) and generate a voltage $\Delta V$. The figure is the schematic of carrier transport in a material with ADCP, which is p-type in $x$ direction and n-type in $y$ direction. (a) and (b) depict the thermoelectric transport in the $x$ and $y$ directions, respectively, for a material that is p-type in the $x$ direction and n-type in the $y$ direction. The thermopower $S = - \Delta V / \Delta T$ is positive (negative) in (a(b)), as the material has hole (electron)-dominated transport in $x(y)$ direction.
    }
    \label{fig: thermopower goniopolar}
\end{figure}

The thermopower of a material measures its ability to generate a voltage difference from a temperature gradient, and is quantitatively characterized by the thermoelectric coefficient:
\begin{equation}
\label{eq:seebeck}
S_{ij} = - \frac{(\Delta V)_i}{(\Delta T)_j},
\end{equation}
where $(\Delta V)_i$ is the difference in voltage along direction $i = x,y,z$ and $(\Delta T)_j$ is the difference in temperature along the direction $j$.
When the generated voltage is along the same direction as the applied temperature gradient, the phenomenon is known as the Seebeck effect, and the corresponding coefficient, $S_{ii}$, is referred to as the Seebeck coefficient. 
The sign of the Seebeck coefficient indicates the polarity of the charge carriers, revealing whether they are n-type or p-type.
In most electronic materials, this polarity is isotropic across different crystallographic directions.
Remarkably, recent experiments (see Ref.~\cite{tianzesong_field_2025} for a review) have discovered that in materials with layered crystal structures, the polarity can be anisotropic, as illustrated in  Fig.~\ref{fig: thermopower goniopolar}.
This type of transport property in materials is called Axis Dependent Conduction Polarity (ADCP), or ``goniopolarity'', meaning that the polarity of the dominant charge carrier depends on the direction of the current.

The discovery of ADCP materials opens new opportunities for a range of applications, particularly related to the thermoelectric effect, that traditionally rely on junctions between two different materials. These applications include transverse thermoelectric generators \cite{zhou_2013} or thermoelectric heaters and coolers without a heterojunction \cite{uchida_thermoelectrics_2022}. In general, ADCP materials enable one to design a point of transition between p-type and n-type transport without heterodoping or an interface between two different materials; one need only change the direction of electric current relative to the crystal axes. Exploring the implications and potential uses of ADCP remains an active topic of study.

Following the initial discovery of ADCP~\cite{he_nasn2as2_2019}, an increasing number of materials have been experimentally identified to manifest ADCP~\cite{tianzesong_field_2025,mg3bi2_goto_band_2024,ochs_nasnas_2021,pdse2_nelson_axis_2023, re4si7_goniopolar, wsi2_ohsumi_transverse_2024,he_nasn2as2_2019,ong_pdcoo2_2010, kmgbi, rowe_thermopower_1970, MnAs_goniopolar}. Nonetheless, there is still no comprehensive theoretical understanding of the conditions that are necessary and sufficient to produce ADCP, in terms of chemical composition or band structure. 
To accelerate the discovery of materials with ADCP and to provide design principles for their realization, in this work we investigate the general conditions for ADCP in both metallic and semiconducting systems.

As the first study aiming to establish such general conditions, we focus on non-interacting systems with relevant low-energy bands having quadratic dispersion. 
We demonstrate that  (i) ADCP can only occur in systems lacking rotational symmetry higher than two-fold; (ii) a proper combination of low-energy electron and hole modes of carrier transport gives rise to ADCP; and (iii) saddle points in the electronic bands are generic sources of ADCP when the Fermi level is sufficiently close.
We derive precise inequalities, in terms of mobility, effective mass, density of states (DoS), and relaxation parameters of the carriers, for the conditions that give rise to ADCP. We validate these inequalities by showing that they are consistent with all known materials with ADCP for which the relevant parameter values are known. In this way our results empower the future discovery of materials with ADCP.

\section{Brief review of Seebeck Coefficient}
As mentioned in the Introduction, the Seebeck coefficient is the longitudinal component of the thermoelectric tensor $\hat{S}$. It can be alternatively defined by considering the equations that govern the electrical and heat current densities, $\mathbf{J}^E$ and $\mathbf{J}^Q$:
\begin{align}
    \mathbf{J}^E &= \hat{\sigma}\mathbf{E} - \hat{\alpha}\grad{T}
    \label{eq: je}
    \\
    \mathbf{J}^Q &= T\hat{\alpha}\mathbf{E} - \hat{\kappa}\grad{T}.
    \label{eq: jq}
\end{align}
Here, $\mathbf{E}$ is the electric field and $\hat{\sigma}, \hat{\alpha}, \hat{\kappa}$ are the electrical, Peltier, and thermal conductivity tensors respectively. The appearance of the same coefficient $\hat{\alpha}$ in the ``off-diagonal'' term of both Eqs.~(\ref{eq: je}-\ref{eq: jq}) is a reflection of Onsager reciprocity. The thermoelectric tensor is defined by $\mathbf{E} = \hat{S} \grad{T}$ under conditions where the electrical current $\mathbf{J}^E = 0$, and therefore
\begin{align}
    \hat{S} = \hat{\sigma}^{-1}\hat{\alpha}.
    \label{generalS}
\end{align}
The electrical and Peltier conductivity tensors can be calculated via the usual Boltzmann equation approach \cite{Ashcroft}:
\begin{align}
    \sigma_{ij} &= \int d\mathcal{E} \left( -\frac{\partial f}{\partial \mathcal{E}} \right) \sigma_{ij} (\mathcal{E}) 
    \label{eq: sigma conductivity}
    \\
    \alpha_{ij} &= \frac{1}{eT} \int d\mathcal{E} (\mathcal{E} - \mu) \left( -\frac{\partial f}{\partial \mathcal{E}} \right) \sigma_{ij} (\mathcal{E}) 
    \label{eq: alpha conductivity}.
\end{align}
Here, $\sigma_{ij}(\mathcal{E})$ corresponds to the conductivity of carriers at energy $\mathcal{E}$ (i.e., to the zero-temperature conductivity when the chemical potential $\mu$ is equal to $\mathcal{E}$) and $f(\mathcal{E}) \equiv \left( \exp\left(\frac{\mathcal{E} - \mu}{k_B T}\right) +1 \right)^{-1}$ is the Fermi-Dirac distribution.

For metals, at low temperatures $(k_B T \ll \mathcal{E}_F) $, calculating the Seebeck coefficients [Eq.~(\ref{generalS})] using the  Sommerfeld expansion in Eqs.~(\ref{eq: sigma conductivity}-\ref{eq: alpha conductivity}), under the assumption of the off-diagonal components of the conductivity tensor being zero, gives us the Mott formula for thermopower \cite{Ashcroft}:
\begin{equation}
    S_{ii} = - \frac{\pi^2 k_B^2 T}{3 {|e|}} \sigma_{ii}^{-1} \frac{d\sigma_{ii}}{d\mathcal{E}}\bigg|_{\mathcal{E}=\mathcal{E}_F},
    \label{eq: Mott formula}
\end{equation}
where $-|e|$ is the electron charge, $\mathcal{E}$ is quasiparticle energy and $\mathcal{E}_F$ is the Fermi energy.

\section{Results}
By definition, a material exhibiting ADCP has different polarities depending on the crystal direction.
Defining the polarity by the sign of the Seebeck effect gives the definition that a material exhibiting ADCP is one for which
\begin{equation}
\label{eq:goniopolar}
S_{ii} S_{jj} < 0
\end{equation}
for some directions $i,j \in {x,y,z}$.
Our analysis in this paper translates this phenomenological condition into constraints on fundamental, experimentally relevant properties of the system, such as symmetries, carrier mobilities and carrier DoS.

\subsection{Symmetry analysis}
Given a crystalline symmetry group $G$ of the system, each symmetry operation $g \in G$ imposes a constraint on the thermoelectric tensor:

\begin{equation}
\label{eq:symmetryconstr}
O_g S O_g^{-1} = S,
\end{equation}
where $O_g$ is the matrix representation of the symmetry operation $g$ in real space. 
It is straightforward to see that reflections and inversion, which have diagonal matrix representations, do not impose constraints on the diagonal components $S_{ii}$. 
However, an $n$-fold (proper or improper) rotation axis with $n>2$ along the $\hat{i} \times \hat{j}$ direction enforces $S_{ii} = S_{jj}$. 
This can be seen most clearly by noting that the quantity $(S_{ii} - S_{jj}, S_{ij} + S_{ji})$ transforms as a vector under a rotation along $\hat{i}\times\hat{j}$ direction. Specifically, a rotation along the $\hat{i}\times\hat{j}$ direction by an angle $2\pi/n$ for a vector in real space rotates the vector  $(S_{ii} - S_{jj}, S_{ij} + S_{ji})$  by $2 \times \frac{2\pi}{n}$, and this vector should be invariant if the rotation is a symmetry of the system according to Eq.~(\ref{eq:symmetryconstr}).
For $n>2$, the only way for this vector to remain invariant under such a rotation is for it to vanish, implying $S_{ii} = S_{jj}$ and $S_{ij} = -S_{ji}$. 
Therefore, to have ADCP in the $\hat{i}$-$\hat{j}$ plane, the system must not have any rotation axis with $n>2$ along $\hat{i}\times\hat{j}$ direction. A detailed analysis is given in Appendix \ref{appendix: symmetry}.

This simple argument already provides a powerful result: ADCP cannot occur in the plane of any crystal that has more than two-fold rotation symmetry. For this reason the majority of known ADCP materials occur in layered compounds, with the layering direction having different polarity than the in-plane directions \cite{tianzesong_field_2025}.

\subsection{ADCP in metals with both electron and hole pockets}
\begin{figure}[htb]
    \centering
    \includegraphics[width=1\columnwidth]{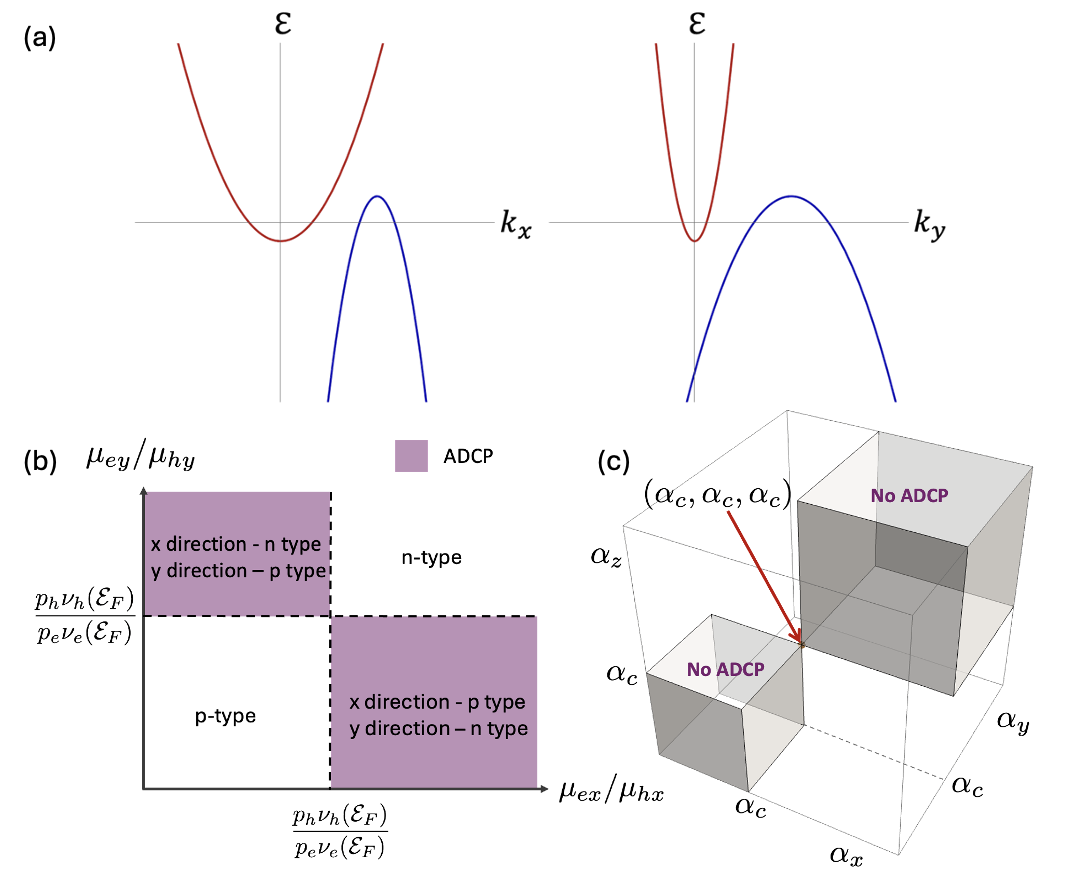}
    \caption{(a) Schematic illustration of the dispersion of a metal with anisotropic electron and hole pockets in the $\mathcal{E}-k_x$ and $\mathcal{E}-k_y$ planes, respectively. (b) Phase diagram of ADCP in 2D materials. $\mu_{e/h,i}$ is the mobility of the electron/hole band in $i$ direction. $p_{e/h}$ captures the energy dependence of the relaxation time [c.f. Eq.~\eqref{eq:relaxationtime}], and $\nu_{e/h}(\mathcal{E}_F)$ is the density of states (DoS) at the Fermi energy $\mathcal{E}_F$. (c) Phase diagram of ADCP in 3D ADCP materials. The $i$-th axis (with $i=x,y,z$) represents a quantity that captures the mobility anisotropies between holes and electrons, $\alpha_i = \mu_{ei}/\mu_{hi}$. The boundaries separating the ADCP and non-ADCP regions are determined by $\alpha_C = p_h \nu_{h}(\mathcal{E}_F)/p_e\nu_{e}(\mathcal{E}_F)$, which depends on relaxation parameters $p_{e/h}$ and $\nu_{e,h}(\mathcal{E}_F)$.}
    \label{fig: 2D Goniopolar}
\end{figure}

Having discussed the symmetry requirement for ADCP, we now consider the simplest situation that gives rise to ADCP: a (semi)metal in which both electron-type and hole-type carrier pockets coexist at the Fermi level.
As illustrated in Fig.~\ref{fig: 2D Goniopolar}(a), if the electrons and holes in such a metal have anisotropic effective masses, then the transport along the direction where electrons (holes) are lighter tends to be $n$-type ($p$-type), which can result into the metal showing ADCP if some conditions involving band structure and relaxation parameters are satisfied. 

Let us start with the case in two dimensions (2D),  where we have one hole pocket and one electron pocket, and try to establish the condition for the presence of ADCP quantitatively.

We use the Mott Formula (Eq.~\ref{eq: Mott formula}) to calculate the Seebeck coefficients. The longitudinal electric conductivity $\sigma_{ii}$ can be calculated using the Drude formula:
\begin{equation}
 \sigma_{ii} = \frac{n_e e^2 \tau_e}{m_{e,i}} + \frac{n_h e^2 \tau_h}{m_{h,i}}. 
\end{equation}
Here, $n_{e(h)}$ are the densities of electrons (holes), $\tau_{e(h)}$ are the relaxation times of electrons (holes), and $m_{e(h),i} = \hbar^2/(d^2 \mathcal{E}_{e(h)}/dk_i^2)$ are the effective masses of electrons (holes) in the $i$ direction $(i = x,y,z)$.

We consider the case where the carriers are electrons and holes in electron and hole pockets described by a quadratic dispersion:
\begin{align}
    \mathcal{E}_{e} &= \mathcal{E}_{e0} + \frac{\hbar^2 k_x^2}{2m_{ex}}  + \frac{\hbar^2 k_y^2}{2m_{ey}}; \\
    \mathcal{E}_{h} &= \mathcal{E}_{h0} - \frac{\hbar^2 k_x^2}{2m_{hx}}  - \frac{\hbar^2 k_y^2}{2m_{hy}};
\end{align}
where $m_{e,x/y} >0$ and $m_{h,x/y} > 0 $ (i.e. the Fermi surfaces are always elliptical). Then, the electron and hole densities at a given Fermi energy $\mathcal{E}_F$ are
\begin{align}
\label{eq: 2d carrier conc}
    n_{e/h} &= \frac{1}{2\pi} \frac{|\mathcal{E}_F-\mathcal{E}_{e/h,0}|}{\hbar^2}\sqrt{m_{e/h,x}m_{e/h,y}} \\ &\equiv \frac{1}{2\pi} \frac{|\mathcal{E}_F-\mathcal{E}_{e/h,0}|}{\hbar^2} m_{e/h},
\end{align}
where we define $m_{e/h}=\sqrt{m_{e/h,x}m_{e/h,y}}$.
Let us further assume that the relaxation times have a power-law dependence on energy:
\begin{align}
\tau_e &= \tau_{e0} |\mathcal{E}_F-\mathcal{E}_{e,0}|^{p_e -1} \\
\tau_h &= \tau_{h0} |\mathcal{E}_F-\mathcal{E}_{h,0}|^{p_h -1},
\end{align}
where $p_h$ and $p_e$ depend on the particular scattering mechanism for momentum-relaxation of carriers in the specific material, such as electron-electron, electron-phonon, and electron-impurity scattering, etc \cite{ECHENIQUE20001_scattering, feliciano_2017, chattopadhyay_1981}. For example, relaxation from impurities with short range potential (long-range Coulomb potential) in a 2D metal with constant DoS corresponds to $p = 1$ ($p=3$).

Given that the conductivity $\sigma_{ii}$ is always positive, the sign of the Seebeck coefficient is determined by the derivative of conductivity with respect to energy:
\begin{align}
    \text{sgn}\;S_{xx} &= -\text{sgn} \left( \frac{d\sigma_{xx}}{d\mathcal{E}_F } \right) 
    \\
    &= -\text{sgn} \left( p_e \mu_{ex}m_e - p_h \mu_{hx}m_h \right), \\
    \text{sgn}\;S_{yy} &= -\text{sgn} \left( \frac{d\sigma_{yy}}{d\mathcal{E}_F}\right) 
    \\
    &= -\text{sgn} \left( p_e \mu_{ey}m_e - p_h \mu_{hy}m_h \right),
\end{align}
where $\mu_{ei} = e\tau_e/m_{ei}$ and $\mu_{hi} = e\tau_h/m_{hi}$ are the mobilities in the $i$-direction for electrons and holes, respectively. 

Note that the effective mass of the electron (hole) with quadratic dispersion is related to the DoS through 
\begin{equation}
    \nu_{e/h}(\mathcal{E}) = \frac{1}{\sqrt{2}\pi^2\hbar^3}m_{e/h}. 
\end{equation}
As a result,  in a rotation-asymmetric 2D material with electron and hole pockets, ADCP [see Eq.~\eqref{eq:goniopolar}] can happen when
\begin{equation}
\label{eq: main result 2d metals}
   \left( \frac{\mu_{ex}}{\mu_{hx}} -\frac{p_h}{p_e}\frac{\nu_h(\mathcal{E}_F)}{\nu_e(\mathcal{E}_F)} \right)\left(\frac{\mu_{ey}}{\mu_{hy}} -\frac{p_h}{p_e}\frac{\nu_h(\mathcal{E}_F)}{\nu_e(\mathcal{E}_F)}\right)<0,
\end{equation}
as illustrated in
Fig.~\ref{fig: 2D Goniopolar}(b). We note that this condition is derived using Eq.~(\ref{eq: 2d carrier conc}), which assumes a parabolic dispersion and therefore is only quantitatively true for parabolic bands. Perturbations to the parabolic dispersions, such as $k^3$ and $k^4$ terms, generally lead to corrections to this condition. These corrections can be neglected when the pocket is sufficiently small, i.e., when the Fermi energy lies sufficiently close to the band extrema, because the quadratic dispersion always dominates in the vicinity of the band extrema. 

A similar process can be extended to three-dimensional (3D) metals, where the carrier density for electron/hole pockets with quadratic dispersion can be expressed as
\begin{equation}
    n_{e/h} = \frac{\sqrt{2}}{3\pi^2\hbar^3}|\mathcal{E}_F - \mathcal{E}_{e/h,0}|^{3/2} \sqrt{\prod_{i = x,y,z} m_{e/h,i}}.
\end{equation}
We also redefine the energy dependence of relaxation times for the 3D case as:
\begin{align}
\label{eq:relaxationtime}
\tau_e &= \tau_{e0} |\mathcal{E}_F-\mathcal{E}_{e,0}|^{p_e -3/2} \\
\tau_h &= \tau_{h0} |\mathcal{E}_F-\mathcal{E}_{h,0}|^{p_h -3/2},
\end{align}
where $p_e$ and $p_h$ again capture the energy dependence of the relaxation times. This redefinition ensures that, in a 3D metal with a density of states $\propto \sqrt{\mathcal{E}}$, impurity scattering from a short-range potential (long-range Coulomb potential) corresponds to $p=1$ ($p=3$), the same as in the 2D case.
We again express the sign of the Seebeck coefficients in terms of the density of states $\nu_{e/h} (\mathcal{E}_F) \propto |\mathcal{E}_F -\mathcal{E}_{e/h,0}|^{1/2} \sqrt{m_{e/h,x} m_{e/h,y}m_{e/h,z} }$ together with the mobilities $\mu_{ei} = e\tau_e/m_{ei}$ and $\mu_{hi} = e\tau_h/m_{hi}$:
\begin{align}
    \text{sgn} (S_{xx}) &= -\text{sgn}\:( p_e \mu_{ex} \nu_{e}(\mathcal{E}_F) - p_h \mu_{hx}\nu_{h}(\mathcal{E}_F)), \\
    \text{sgn} (S_{yy}) &= -\text{sgn}\:( p_e \mu_{ey} \nu_{e}(\mathcal{E}_F) - p_h \mu_{hy}\nu_{h}(\mathcal{E}_F)), \\
    \text{sgn} (S_{zz}) &= -\text{sgn}\:( p_e \mu_{ez} \nu_{e}(\mathcal{E}_F) - p_h \mu_{hz}\nu_{h}(\mathcal{E}_F)).
\end{align}

As a result, the condition for ADCP is given by:
\begin{equation}
    \label{eq: main condition e h metal}
   \left( \frac{\mu_{ei}}{\mu_{hi}} -\frac{p_h}{p_e}\frac{\nu_h(\mathcal{E}_F)}{\nu_e(\mathcal{E}_F)} \right)\left(\frac{\mu_{ej}}{\mu_{hj}} -\frac{p_h}{p_e}\frac{\nu_h(\mathcal{E}_F)}{\nu_e(\mathcal{E}_F)}\right)<0,
\end{equation}
for some choice of directions $i, j$. Explicitly, this inequality says that if there exist two directions $i$ and $j$ in a material such that $\alpha_i = \mu_{ei}/\mu_{hi}$ is bigger than the cutoff value $\alpha_c = p_h\nu_h(\mathcal{E}_F)/p_e\nu_e(\mathcal{E}_F)$, and $\alpha_j= \mu_{ej}/\mu_{hj}$ is smaller than the cutoff value, then the material exhibits ADCP in the $\hat{i}-\hat{j}$ plane, as shown in Fig.~\ref{fig: 2D Goniopolar}.

The conclusion for the simplest one-electron and one-hole pocket case can be easily extended to the general case with many carrier pockets. Within our assumptions, the material will exhibit ADCP in the $\hat{i}-\hat{j}$ plane (i.e.\ will have different transport polarities in the $i$ and $j$ directions) if and only if
\begin{equation}
\begin{aligned}
   &\left( \frac{\sum_{s}^{N_e}\mu_{ei,s}\nu_{e,s}(\mathcal{E}_F)}{\sum_{s'}^{N_h} \mu_{hi,s'}\nu_{h,s'}(\mathcal{E}_F)} - \frac{p_h }{p_e } \right) 
   \\ &\times \left(   \frac{\sum_{s}^{N_e}\mu_{ei,s}\nu_{e,s}(\mathcal{E}_F)}{\sum_{s'}^{N_h} \mu_{hi,s'}\nu_{h,s'}(\mathcal{E}_F)} - \frac{p_h }{p_e } \right) < 0,
\end{aligned}
\end{equation}
where $s(s')$ are band indices for electron (hole) band and $N_{e(h)}$ is the total number of electron (hole) pockets.

\subsection{Saddle points as sources for ADCP}
    \label{sec: saddle pt calc}

\begin{figure}
    \centering
    \includegraphics[width=1\columnwidth]{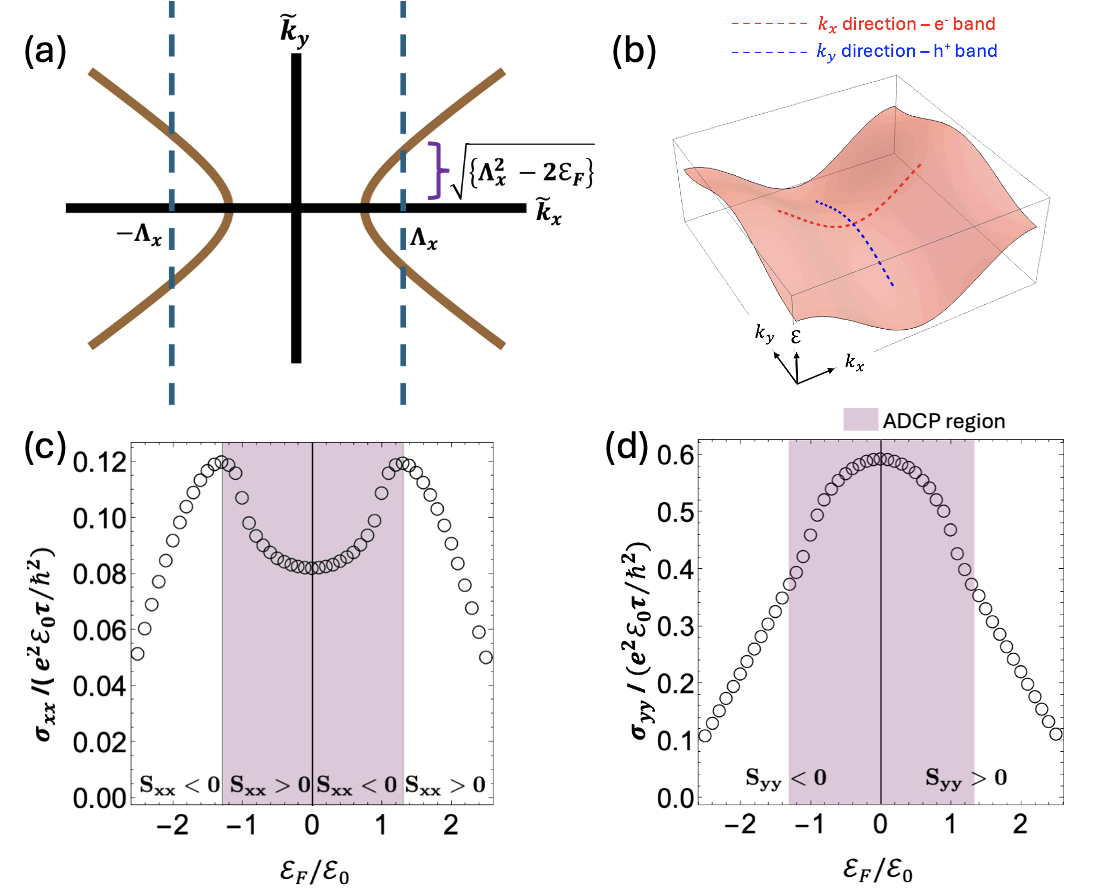}
    \caption{(a) Illustration of a Fermi surface near a saddle point.  $\Lambda_x$ is a cutoff for $\Tilde{k}_x$. (b) Schematic of a saddle point dispersion (in pink). The Fermi surface has both convex and concave parts and therefore gives rise to both electron (red dashed line) and hole (blue dashed line) dispersion.  (c \& d) Longitudinal conductivities in $x$ and $y$ direction as a function of Fermi energy. The purple region shows the region (Fermi energies) where ADCP is seen. In these regions, the conductivities in $x$ (b) and $y$ (c) directions have derivatives of different signs, and therefore the Seebeck coefficients have different signs. 
    Note that the conductivity plotted here is expressed in units of $e^2 \mathcal{E}_0 \tau / \hbar^2$ rather than the conventional unit $e^2/\hbar$. Since the momentum relaxation time $\tau$ in a metal is typically much larger than $\hbar/\mathcal{E}_0$, the apparently small conductivity values in the plots actually correspond to very large conductivities in units of $e^2/\hbar$, as expected.
    } 
    \label{fig: saddlepoint fig}
\end{figure}

So far we have considered systems with multiple bands that coexist at the Fermi level. We now turn our attention to materials with a single Fermi surface, motivated by the first experimental discovery of ADCP in NaSn\textsubscript{2}As\textsubscript{2} \cite{he_nasn2as2_2019}, which has a single Fermi surface with convex and concave parts.
In order for such a material to show ADCP, the surface should be electron-like in one direction and hole-like in the perpendicular direction. 
Therefore, we consider a Fermi surface around a saddle point, as depicted in Fig.~\ref{fig: saddlepoint fig}. Such saddle points are common in electronic bands. 
By explicitly calculating the sign of the Seebeck coefficients in different directions, we show in this section that saddle points are generic sources of ADCP. 

We study the Fermi surface near a saddle point with the generic dispersion
\begin{equation}
\label{eq: saddle_dispersion}
    \mathcal{E}_{\mathbf{k}} = \frac{\hbar^2 k_x^2}{2m_x} -  \frac{\hbar^2 k_y^2}{2m_y},
\end{equation}
where $m_x \neq m_y$.  
We use the Mott formula (Eq.~(\ref{eq: Mott formula})) to calculate the sign of the Seebeck coefficient. For a saddle point dispersion depscribed by Eq.~(\ref{eq: saddle_dispersion}), the conductivity (Eq.~(\ref{eq: sigma conductivity}) under the relaxation time approximation becomes:
\begin{equation}
    \sigma_{ii} = \frac{e^2 \tau}{(2\pi)^2} \int d^2k \, v_{F,i}^2(\mathbf{k}) \delta \left( \frac{\tilde{k}_x^2}{2} - \frac{\tilde{k}_y^2}{2} - \mathcal{E}_F \right),
    \label{eq: saddle pt conductivity}
\end{equation}
where $\Tilde{k}_i = \hbar k_i/ \sqrt{m_i}$, and the Fermi energy $\mathcal{E}_{F}$ is measured relative to the saddle point at $\mathcal{E}=0$.  The details of the derivation of Eq.~(\ref{eq: saddle pt conductivity}) can be found in Appendix~\ref{appendix: saddlept}.

Eq.~\eqref{eq: saddle_dispersion} is an effective low-energy description around a high-symmetry momentum, and becomes invalid at some momentum scale $\sqrt{m_i}\Lambda_x/\hbar$, where $\Lambda_x $ is the cutoff for $\tilde{k}_x$ and depends on the details of the band structure, as illustrated in Fig.~\ref{fig: saddlepoint fig}.
With this cutoff, we can calculate the longitudinal conductivities and thus their derivatives with respect to energy in both $x$ and $y$ directions, which determines the sign of Seebeck coefficient according to the Mott formula [Eq.~(\ref{eq: Mott formula})]:
\begin{multline}
    \frac{d\sigma_{xx}}{d\mathcal{E}_F} = \frac{4e^2\tau}{(2\pi)^2\hbar^2} \sqrt{\frac{m_y}{m_x}} \left( \log \frac{\Lambda_x + \sqrt{\Lambda_x^2 +2\mathcal{E}_F}}{\sqrt{2\mathcal{E}_F}}\right. \\ \left. - \frac{\Lambda_x}{\sqrt{\Lambda_x^2-2\mathcal{E}_F}} \right), \label{eq: dsigmaxx}
\end{multline}
\begin{multline}
    \frac{d\sigma_{yy}}{d\mathcal{E}_F} = \frac{4e^2\tau}{(2\pi)^2\hbar^2} \sqrt{\frac{m_x}{m_y}} \left( -\log \frac{\Lambda_x + \sqrt{\Lambda_x^2 -2\mathcal{E}_F}}{\sqrt{2\mathcal{E}_F}} \right). \label{eq: dsigmayy}
\end{multline}
The details of this calculation can be found in Appendix \ref{appendix: saddlept}. Note that we assume $\Lambda_x^2 > 2\mathcal{E}_F$, so that a finite Fermi surface exists within the cutoff. 
From Eq.~\eqref{eq: dsigmayy}, we can see that $d\sigma_{yy}/d\mathcal{E} $ is always negative and thus $S_{yy}$ is always positive. 
As a result, ADCP arises when $d\sigma_{xx}/d\mathcal{E}$ is positive.
According to Eq.~\eqref{eq: dsigmaxx}, we can derive that $d\sigma_{xx}/d\mathcal{E}>0 $ when $\Lambda_x^2 > 2.31 \mathcal{E}_F$.

Therefore, at low temperature and when the Fermi surface is close enough to the saddle point (assumed to be at $\mathcal{E}=0$), such that the conditions
\begin{equation}
    \label{eq: main inequality - saddlept}
    k_B T \ll \mathcal{E}_F < \Lambda_x^2/2.31
\end{equation}
are satisfied, ADCP arises generically. 

In order to demonstrate the appearance of ADCP for a single Fermi surface near a saddle point numerically, we consider a simple model with dispersion
\begin{equation}
\label{eq:latticemodel}
    \mathcal{E}_{\textbf{k}} = \mathcal{E}_0(\cos{k_x a} - 2\cos{k_y a}),
\end{equation}
where $\mathcal{E}_0$ is an energy scale capturing the band width and $a$ is the lattice constant.

It is straightforward to see that the band described by Eq.~\eqref{eq:latticemodel} has one minimum at $\mathcal{E}/\mathcal{E}_0=-3$, one maximum at $\mathcal{E}/\mathcal{E}_0=+3$, and two saddle points at different energies -- one at $\mathcal{E}/\mathcal{E}_0=-1$ and the other at $\mathcal{E}/\mathcal{E}_0=+1$.
We can tune the chemical potential to control the distance between the fermi surface and saddle points, and explore the possible ADCP phases. 
In Fig.~\ref{fig: saddlepoint fig}(b-c), we plot the longitudinal conductivities as a function of Fermi energy for this system. As mentioned before, the condition for ADCP is related to the sign of the Seebeck coefficient, which is opposite to the sign of the  derivative of the longitudinal conductivity with respect to the Fermi energy (Eq.~(\ref{eq: Mott formula})). Therefore, in Fig.~\ref{fig: saddlepoint fig}(b-c), the regions with increasing and decreasing conductivity have different signs of the Seebeck coefficient. The purple region is the region where $S_{xx}$ and $S_{yy}$ are of opposite signs and thereby the transport has ADCP. This region ranges from $\mathcal{E}_F/\mathcal{E}_0 \approx -1.3 $ to $\mathcal{E}_F/\mathcal{E}_0 \approx 1.3 $. When $0<\mathcal{E}_F/\mathcal{E}_0 < 1.3$, we have $S_{xx} > 0$ and $S_{yy} < 0$; whereas for $-1.3<\mathcal{E}_F/\mathcal{E}_0 <0$, $S_{xx} < 0$ and $S_{yy} >0$. 

Interestingly, our numerical results show that the ADCP transition happens near the energy associated with the Lifshitz transition, at which the topology of the Fermi surface changes. 
We would like to comment on whether there is a connection between these two transitions as the Fermi level is tuned.
At the Lifshitz transition, the energy derivative of the longitudinal conductivity $\sigma_{ii}$ diverges. 
On the other hand, at the ADCP transition, the energy derivative of $\sigma_{ii}$ changes sign (for one of the directions). 
Since a divergence of this derivative does not necessarily indicate a sign change of it, an ADCP transition does not necessarily happen simultaneously with a Lifshitz transition.  In our simple model, the sign change of energy derivative of $\sigma_{xx}$ actually occurs through the derivative going through a zero. 
The reasons that the two transitions occur around the same value of the Fermi energy are (i) ADCP requires the Fermi surface to be close to a saddle point and (ii) the Lifshitz transition occurs when the Fermi surface passes through a saddle point. How close these two transitions are depends on details of the band structure and is not a universal property.

We also highlight that in real materials with chemical potential near a saddle point, the polarity of thermal transport should be highly tunable— 
the material can be tuned into exhibiting ADCP or having isotropic transport by adjusting the carrier doping.

\subsection{ADCP in intrinsic semiconductors with thermally activated carriers}

\begin{figure}[htb]
    \centering
    \includegraphics[width=0.9\columnwidth]{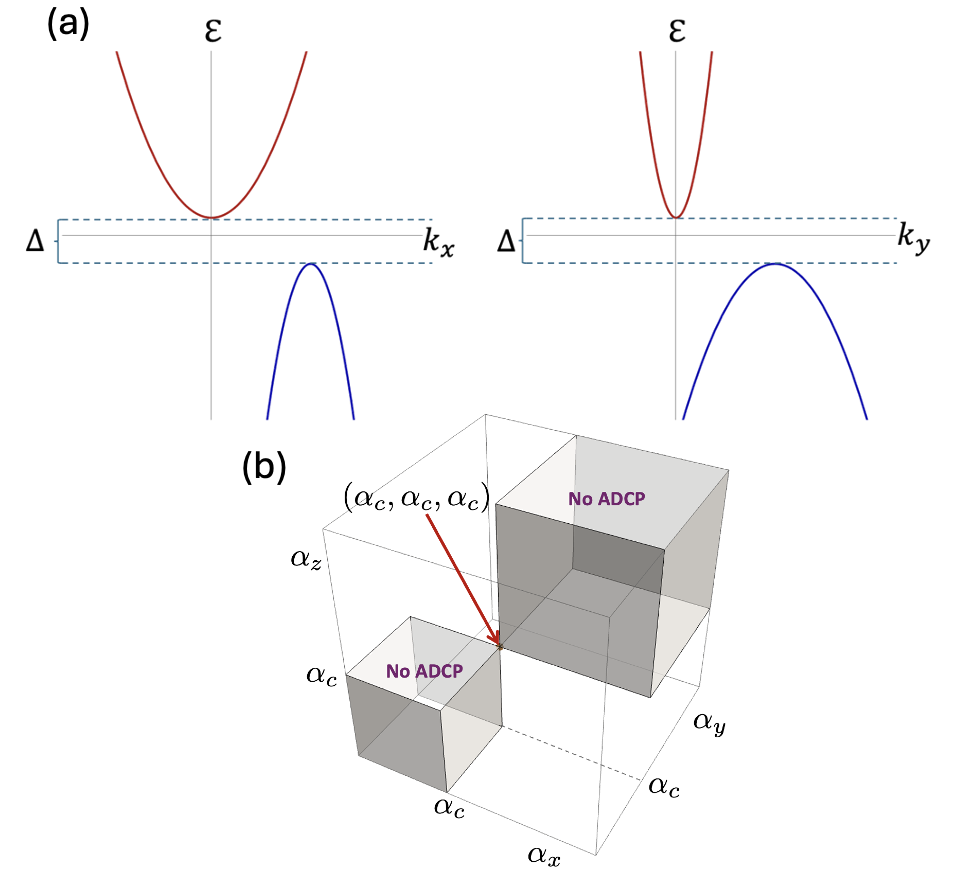}
    \caption{(a) Schematic of the dispersion of a multiband semiconductor with electron and hole pockets, projected to the $\mathcal{E}-k_x$ and $\mathcal{E}-k_y$ planes, respectively. The band gap is $\Delta$. (b) Phase diagram for a 3D semiconductor showing ADCP. The axis has a quantity which depends on the mobility anisotropies of holes and electrons $\alpha_i = m_{hi}/m_{ei}$, and the boundaries $\alpha_C = \tau_h m_h^{3/2}/\tau_e m_e^{3/2}$ that separate ADCP and non-ADCP regions depend on the relaxation parameters $\tau$ and the effective masses $m_{e/h}$.
    }
    \label{fig: insulator}
\end{figure}

After studying conditions for ADCP in metals, let us now consider the case of semiconductors with thermally excited carriers, for which ADCP can arise if the conduction and valence bands are sufficiently anisotropic in different directions (as depicted in Fig.~\ref{fig: insulator}(a)). We focus on the case of intrinsic (undoped) semiconductors with a band gap $\Delta$. 
The Mott formula (Eq.~(\ref{eq: Mott formula})) is not valid in insulators because the Mott formula is derived using the Sommerfeld expansion, which assumes existence of a Fermi surface and low temperature $(k_B T \ll \mathcal{E}_F$). Instead, we use Eq.~(\ref{generalS}) to calculate and analyze the Seebeck coefficient.

We first calculate the Seebeck coefficient in the $x$ direction through $S_{xx} = \alpha_{xx} /\sigma_{xx}$, assuming the off diagonal components of the conductivity (and thus the thermoelectric tensor) to be zero. Here,
\begin{align}
    \alpha_{xx} &= \alpha^e_{xx} + \alpha^h_{xx},
    \label{eq: alpha total}
    \\
    \sigma_{xx} &= \sigma^e_{xx} + \sigma^h_{xx}.
    \label{eq: sigma total}
\end{align}
We consider a parabolic electron dispersion, with different effective masses in different directions:
\begin{equation}
    \mathcal{E} = \frac{1}{2} \left( \frac{\hbar^2k_x^2}{m_{ex}} + \frac{\hbar^2k_y^2}{m_{ey}} + \frac{\hbar^2k_z^2}{m_{ez}} \right).
\end{equation}
We focus on intrinsic semiconductors, in which the electron and hole densities are equal. Under the low-temperature condition $k_B T \ll \Delta$, the chemical potential lies at the middle of the band gap, i.e., $E_c-\mu = \mu-E_v = \Delta/2$, where $E_c$ ($E_v$) indicates the energy of the conduction (valence) band edge, and the carriers are in the nondegenerate limit, meaning that the Fermi--Dirac distribution can be well approximated by the Boltzmann distribution for both bands. With these conditions and the assumption that the relaxation time is energy independent, we derive the Seebeck coefficient (see details in Appendix \ref{conductivity-semiconductor}) as
\begin{align}
    S_{xx} &= -\frac{1}{eT} \frac{A_{ex}\tau_e- A_{hx}\tau_h}{A_{ex}\tau_e+ A_{hx}\tau_h} \left( \frac{5}{2} k_BT + \frac{\Delta}{2} \right), \\
    S_{yy} &= -\frac{1}{eT} \frac{A_{ey}\tau_e- A_{hy}\tau_h}{A_{ey}\tau_e+ A_{hy}\tau_h} \left( \frac{5}{2} k_BT + \frac{\Delta}{2} \right),
\end{align}
where $A_{e(h)x} =  \frac{2\sqrt{2}}{3} \frac{e^2}{\pi^2\hbar^3} \left( \frac{m_{e(h)y}m_{e(h)z}}{m_{e(h)x}} \right)^{1/2} $ and $A_{e(h)y} =  \frac{2\sqrt{2}}{3} \frac{e^2}{\pi^2\hbar^3} \left( \frac{m_{e(h)x}m_{e(h)z}}{m_{e(h)y}} \right)^{1/2} $.
For such a material to show ADCP, we need
\begin{multline}
    S_{xx} > 0 \; \text{and} \; S_{yy} <0 \implies \\ A_{ex}\tau_e <A_{hx}\tau_h \; \text{and} \;  A_{ey}\tau_e >A_{hy}\tau_h \:\: \text{OR} \\
     S_{xx} < 0 \; \text{and} \; S_{yy} >0 \implies \\ A_{ex}\tau_e >A_{hx}\tau_h \; \text{and} \;  A_{ey}\tau_e <A_{hy}\tau_h.
\end{multline}
This condition implies
\begin{align}
    \left(\frac{m_{hx}}{m_{ex}} - \frac{\tau_h m_h^{3/2}}{\tau_e m_e^{3/2}} \right) \left( \frac{m_{hy}}{m_{ey}} -\frac{\tau_h m_h^{3/2}}{\tau_e m_e^{3/2}} \right) < 0.
\end{align}
More generally, the condition for ADCP to emerge in the $\hat{i}-\hat{j}$ plane can be expressed as
\begin{align}
    \label{eq: main result semiconductor goniopolarity condition}
    \left(\frac{m_{hi}}{m_{ei}} - \frac{\tau_h m_h^{3/2}}{\tau_e m_e^{3/2}} \right) \left( \frac{m_{hj}}{m_{ej}} -\frac{\tau_h m_h^{3/2}}{\tau_e m_e^{3/2}} \right) < 0.
\end{align}
This condition is illustrated in Fig.~\ref{fig: insulator}(b).

We also derive the condition for ADCP in intrinsic semiconductors with thermally activated carriers that have energy-dependent relaxation times. These inequalities are dependent on the band gap $\Delta$ and the temperature $T$, in addition to the usual band and relaxation parameters. We have discussed the derivation and the conditions in Appendix \ref{appendix: energy dependent relaxation times}.

\section{Conclusion and remarks on materials}

In this article, we have derived conditions for 2D and 3D metallic and semiconducting systems with electron and hole pockets to exhibit ADCP. These conditions are inequalities featuring effective masses, mobilities and relaxation parameters, which the material needs to satisfy to exhibit ADCP. These inequalities constitute our main results and are given by Eqs.(\ref{eq: main result 2d metals}, \ref{eq: main condition e h metal},\ref{eq: main result semiconductor goniopolarity condition}).

We have also studied metals  
with a single Fermi surface for which the Fermi energy is near a saddle point. We analytically demonstrate that the system manifests ADCP whenever the Fermi energy is sufficiently close enough to the saddle point, i.e., $(\mathcal{E}_F-\mathcal{E}_{\text{saddle}})< \hbar^2 (k\text{-cutoff})^2/2.31 m$, with $k\text{-cutoff}$ determined by the details of the band structure. 

We can validate the ADCP criteria we have developed by checking whether they are satisfied in known ADCP materials. To this end
we present in Table \ref{table: goniopolar materials} a summary of known ADCP materials, their mechanism for ADCP (multicarrier or  single Fermi surface) and comment on whether the appearance of ADCP agrees with our theory.

\begin{table*}[!ht]
\centering
 \begin{tabular}{||c c c c c||} 
 \hline
 \textbf{ADCP Material} & \textbf{Type} & \textbf{Reference} &  \textbf{\makecell{Relevant \\ Inequality}} & \textbf{\makecell{Do the parameters \\ satisfy the inequality?}} \\ [0.2ex] 
 \hline\hline
 CsBi\textsubscript{4}Te\textsubscript{6} & Semiconductor & \cite{chung_anisotropy_csbi4te6_2003} & Eq.~(\ref{eq: main result semiconductor goniopolarity condition}) & Yes \\ \hline
  Mg\textsubscript{3}Sb\textsubscript{2} & Semiconductor & \cite{mg3bi2_goto_band_2024} & Eq.~(\ref{eq: main result semiconductor goniopolarity condition}) & Yes \\ \hline
 NaSnAs & Semiconductor & \cite{ochs_nasnas_2021} & Eq.~(\ref{eq: main result semiconductor goniopolarity condition}) & Yes\\ \hline
 PdSe\textsubscript{2} & Semiconductor & \cite{pdse2_nelson_axis_2023} & Eq.~(\ref{eq: main result semiconductor goniopolarity condition}) & Yes\\ \hline
 Re\textsubscript{4}Si\textsubscript{7} & Semiconductor & \cite{re4si7_goniopolar} & Eq.~(\ref{eq: main result semiconductor goniopolarity condition}) & Yes\\ \hline
 WSi\textsubscript{2} & Semiconductor & \cite{wsi2_ohsumi_transverse_2024} & Eq.~(\ref{eq: main result semiconductor goniopolarity condition}) & Yes\\ \hline
  Mg\textsubscript{3}Bi\textsubscript{2} & Multi Fermi Surface Metal & \cite{mg3bi2_goto_band_2024} & Eq.~(\ref{eq: main result semiconductor goniopolarity condition})  & Yes\\ \hline
  KMgBi & Multi Fermi Surface Metal & \cite{kmgbi} & Eq.~(\ref{eq: main result semiconductor goniopolarity condition}) & Data Unavailable \\ \hline
  ZrTe\textsubscript{3} & 2D Semiconductor & \cite{felser_electronic_1998_zrte3}& Eq.~(\ref{eq: main result semiconductor goniopolarity condition}) & Data Unavailable \\ \hline
 NaSn\textsubscript{2}As\textsubscript{2} & Single Fermi Surface Metal & \cite{he_nasn2as2_2019} & Eq.~(\ref{eq: main inequality - saddlept}) & Data Unavailable \\ \hline
 PdCoO\textsubscript{2} & Single Fermi Surface Metal & \cite{ong_pdcoo2_2010} & Eq.~(\ref{eq: main inequality - saddlept}) &Data Unavailable\\ \hline
 LaPt\textsubscript{2}B & Single Fermi Surface Metal & \cite{manako_large_2024_lapt2b} & Eq.~(\ref{eq: main inequality - saddlept}) &Data Unavailable \\ \hline
 \end{tabular}
 \label{table: goniopolar materials}
 \caption{List of materials experimentally found to show ADCP. For each material, we mention the relevant result that we have derived and report whether the band and relaxation parameters satisfy the relevant inequality.}
\end{table*}

The ADCP materials CsBi\textsubscript{4}Te\textsubscript{6}\cite{chung_anisotropy_csbi4te6_2003}, Mg\textsubscript{3}Sb\textsubscript{2}\cite{mg3bi2_goto_band_2024}, NaSnAs\cite{ochs_nasnas_2021}, PdSe\textsubscript{2}\cite{pdse2_nelson_axis_2023}, Re\textsubscript{4}Si\textsubscript{7}\cite{re4si7_goniopolar} and WSi\textsubscript{2}\cite{wsi2_ohsumi_transverse_2024} are semiconductors. We use their effective masses to verify that they satisfy the condition for ADCP given by Eq.(\ref{eq: main result semiconductor goniopolarity condition}). 
Note that since the relaxation time ratio between holes and electrons for these materials is not reported in literature, we have assumed this ratio to be unity for the purpose of our estimates, because this ratio is typically in the order of unity in most semiconductors. Mg\textsubscript{3}Bi\textsubscript{2}\cite{mg3bi2_goto_band_2024} is a topological semimetal, but it shows ADCP through the multicarrier (electron and hole pockets separated by a gap) mechanism. Thus,  we also used the reported effective masses and mobilities to verify that this material satisfies Eq.(\ref{eq: main result semiconductor goniopolarity condition}). 

The ADCP materials NaSn\textsubscript{2}As\textsubscript{2}\cite{he_nasn2as2_2019} and  
PdCoO\textsubscript{2}\cite{ong_pdcoo2_2010} are metals with a single Fermi surface near a saddle point.  However, due to their complicated Fermi surfaces provided in the literature, it is difficult to identify the exact k-space coordinates and the energy for the saddle points, which requires further detailed Density Functional Theory (DFT) calculations. 
To predict whether such a material with a saddle point will manifest ADCP, the details needed are the energy difference between the Fermi energy and the saddle point energy, and the band dispersion around the saddle point. Specifically, according to Sec.\ref{sec: saddle pt calc}, if the momentum cutoff for the approximate expression Eq.(\ref{eq: saddle pt conductivity}) and the difference between the Fermi energy and the energy of the saddle point satisfy the condition $(\mathcal{E}_F-\mathcal{E}_{\text{saddle}})< \hbar^2 (k\text{-cutoff})^2/2.31 m$, ADCP emerges. The cutoff $k$-cutoff in this inequality can be estimated using details of the dispersion from DFT calculations.

Other ADCP materials known presently are LaPt\textsubscript{2}B\cite{manako_large_2024_lapt2b}, KMgBi\cite{kmgbi} and ZrTe\textsubscript{3}\cite{felser_electronic_1998_zrte3}. We were unable to verify our theory for KMgBi\cite{kmgbi} and ZrTe\textsubscript{3}\cite{felser_electronic_1998_zrte3} due to the lack of effective mass data. LaPt\textsubscript{2}B\cite{manako_large_2024_lapt2b} contains mixed dimension Fermi surfaces, a case which is outside the scope of our theory.

We conclude by noting that our concrete criteria motivate exploring situations where ADCP can be controllably switched on and off in well-studied materials like silicon, III-V semiconductors (GaAs), $\alpha$-tin, etc. For example, it is promising to engineer strain to get ADCP in Luttinger-Kohn semiconductors. As demonstrated in Ref.~\cite{sun_strain_2010}, strain affects electron and hole bands differently, and will lead to strong anisotropy in one of them but not the other. In this way one can envision turning ADCP on and off in certain materials by the application of strain.

\section*{Acknowledgements}

The authors thank Shaffique Adam and Tianze Song for helpful discussions.
This work was supported by the Center for Emergent Materials, an NSF-funded MRSEC, under Grant No.\ DMR-2011876.

\appendix
\begin{widetext}
\section{Symmetry analysis -- cannot have rotation axis with $n>2$.}
\label{appendix: symmetry}

\textbf{2D:}

We analyze the constraints an ADCP material should have starting from crystalline point group symmetry. We assume that a material has a point group symmetry $G$ (generally the orthogonal group $O(2)$ or the subgroup $O(2)$). We also assume that any elements $O$ in $G$, the Seebeck coefficient should satisfy:
\begin{equation}
    OSO^T = S.
    \label{eq: seebeck symmetry}
\end{equation}

At first, we consider the constraints from n-fold rotation symmetries that are compatible with translation symmetry (n = 2,3,4,6, i.e. , $\theta = 2\pi /n$ in the following equation):

\begin{align}
\begin{pmatrix}
S_{xx} & S_{xy}\\
S_{yx} & S_{yy}
\end{pmatrix} = \begin{pmatrix}
\cos{\theta} & -\sin{\theta}\\
\sin{\theta} & \cos{\theta}
\end{pmatrix} \begin{pmatrix}
S_{xx} & S_{xy}\\
S_{yx} & S_{yy}
\end{pmatrix} \begin{pmatrix}
\cos{\theta} & \sin{\theta}\\
-\sin{\theta} & \cos{\theta}
\end{pmatrix}
\end{align}

From these equations, we conclude that if there exists a rotation symmetry with $n >2$, the Seebeck coefficients should satisfy $S_{xx} = S_{yy}$ and $S_{xy} = -S_{yx}$. Reflection symmetry along $x$ and/or $y$ would require $S_{xy} = S_{yx} = 0$. Thus, a two-dimensional ADCP material cannot have n-fold rotation symmetry with $n>2$.

\textbf{3D:}

We assume that a material has a point group symmetry $G$ (generally the orthogonal group $O(3)$ or the subgroup $O(3)$). We also assume that any elements $O$ in $G$, the Seebeck coefficient should satisfy Eq.(\ref{eq: seebeck symmetry}). 

For example, for rotation around $x$ axis, the Seebeck coefficient would satisfy:

\begin{gather}
\begin{pmatrix}
S_{xx} & S_{xy} & S_{xz}\\
S_{yx} & S_{yy} & S_{yz} \\
S_{zx} & S_{zy} & S_{zz}
\end{pmatrix} = \begin{pmatrix}
1 & 0 & 0 \\
0 & \cos{\theta} & -\sin{\theta}\\
0 & \sin{\theta} & \cos{\theta}
\end{pmatrix} \begin{pmatrix}
S_{xx} & S_{xy} & S_{xz}\\
S_{yx} & S_{yy} & S_{yz} \\
S_{zx} & S_{zy} & S_{zz}
\end{pmatrix} \begin{pmatrix}
1 & 0 & 0 \\
0 & \cos{\theta} & \sin{\theta}\\
0 & -\sin{\theta} & \cos{\theta}
\end{pmatrix}
\end{gather}

From this equation, we come to the same conclusion, if there exists a rotation symmetry with $n >2$, the Seebeck coefficients should satisfy $S_{xx} = S_{yy}$ and $S_{xy} = -S_{yx}$. Reflection symmetry along $x$ and/or $y$ would require $S_{xy} = S_{yx} = 0$. Thus, a three-dimensional ADCP material cannot have n-fold rotation symmetry with $n>2$.

A three dimensional ADCP material can have inversion symmetry. For mirror symmetries, Eq.(\ref{eq: seebeck symmetry}), we find that the off-diagonal components of $S$ have to be zero, and therefore this symmetry does not prohibit ADCP either.

\section{Calculation of Conductivities for Saddle Points}\label{appendix: saddlept}

We calculate the longitudinal conductivity of such a material, by calculating the average of $\langle v_i^2\rangle$ locally around the saddle point and introducing a cutoff $\Lambda_x$, i.e.
\begin{equation}
    \sigma_{ii} = e^2 \nu(\mathcal{E})\langle v_i(\mathcal{E}) \rangle^2_{\text{F.S.}} \tau (\mathcal{E})
\end{equation}

Dispersion close to the saddle point can be written as:
\begin{equation}
    \mathcal{E}_{\mathbf{k}} = \frac{\hbar^2 k_x^2}{2m_x} -  \frac{\hbar^2 k_y^2}{2m_y}.
\end{equation}
To simplify the expression, we define variables $\Tilde{k_i} = \hbar k_i / \sqrt{m_i}$ for $i = x,y$, and then the dispersion becomes:
\begin{equation}
    \mathcal{E}_{\mathbf{k}} = \frac{\Tilde{k}_x^2}{2} -  \frac{\Tilde{k}_y^2}{2}.
\end{equation}
Ignoring the spin degree of freedom, we can write the electric current density as
\begin{equation}
    \mathbf{j} = - \frac{e}{(2\pi)^2} \int d^2k \mathbf{v}f_{\mathbf{k}},
\end{equation}
where $f_{\mathbf{k}}$ is the steady state distribution of the carriers.
To calculate the longitudinal conductivity, we calculate
\begin{equation}
    \frac{df_{\mathbf{k}}}{dt} = \frac{\partial f_{\mathbf{k}}}{\partial t} + \dot{\mathbf{k}}\cdot \frac{\partial f_{\mathbf{k}}}{\partial \mathbf{k}} + \dot{\mathbf{r}}\cdot \frac{\partial f_{\mathbf{k}}}{\partial \mathbf{r}} = - \frac{f_{\mathbf{k}}-f_0}{\tau}.
    \label{eq: dfkdt}
\end{equation}
As $f_{\mathbf{k}}$ is the steady state distribution, the $\frac{\partial f_{\mathbf{k}}}{\partial t} = 0$ and $\frac{\partial f_{\mathbf{k}}}{\partial \mathbf{r}} =0$. The last expression is written using the relaxation time approximation and $\tau$ is the relaxation time. $f_0$ is thermal equilibrium distribution of carriers.
Defining a variable $f_1 = f_{\mathbf{k}} - f_0 \ll f_0$, and using the Eq.(\ref{eq: dfkdt}), we get
\begin{align}
-\frac{e}{\hbar}\mathbf{E}\cdot \frac{\partial f_0}{\partial \mathbf{k}} &= - \frac{f_1}{\tau} \\
\implies  e\mathbf{E}\cdot \mathbf{v}_{\mathbf{k}} \frac{\partial f_0}{\partial \mathcal{E}} &= \frac{f_1}{\tau}. 
\end{align}
Therefore
\begin{equation}
    f_1 = e\tau \mathbf{E}\cdot \mathbf{v}_{\mathbf{k}} \frac{\partial f_0}{\partial \mathcal{E}} \approx -  e\tau \mathbf{E}\cdot \mathbf{v}_{\mathbf{k}} \delta\left( \frac{\Tilde{k}_x^2}{2} -  \frac{\Tilde{k}_y^2}{2} - \mathcal{E}_{\mathbf{k}} \right).
\end{equation}
We can write the longitudinal conductivity as 
\begin{equation}
    \sigma_{ii} = \frac{e^2 \tau}{(2\pi)^2} \int d^2k v_{F,i}^2(\mathbf{k}) \delta \left( \frac{\Tilde{k}_x^2}{2} - \frac{\Tilde{k}_y^2}{2} - \mathcal{E}_F \right). 
\end{equation}

We introduce an $x$- momentum $(k_x)$ cutoff $\Lambda_x$, as shown in Fig. \ref{fig: saddlepoint fig}, and then calculate the longitudinal conductivities in $x$ and $y$ directions. 

\begin{align}
    \sigma_{xx} &= \frac{4e^2\tau\sqrt{m_xm_y}}{(2\pi)^2\hbar^2} \int_0^{\Lambda_x} \int_0^{\sqrt{\Lambda_x^2 - 2\mathcal{E}_F}} d\Tilde{k}_x d\Tilde{k}_y 
    \frac{\Tilde{k}_x^2}{m_x}  \delta \left( \frac{\Tilde{k}_x^2}{2} - \frac{\Tilde{k}_y^2}{2} - \mathcal{E}_F \right) \\
     &= \frac{4e^2\tau\sqrt{m_xm_y}}{(2\pi)^2\hbar^2} \int_0^{\sqrt{\Lambda_x^2 - 2\mathcal{E}_F}} d\Tilde{k}_y \frac{\sqrt{\Tilde{k}_y^2 + 2\mathcal{E}_F}}{m_x} \\
     &= \frac{4e^2\tau\sqrt{m_xm_y}}{(2\pi)^2\hbar^2} \frac{1}{2}  \left( \Lambda_x \sqrt{\Lambda_x^2 -2\mathcal{E}_F} + 2\mathcal{E}_F \log \frac{\sqrt{\Lambda_x^2 -2\mathcal{E}_F}+\Lambda_x}{\sqrt{2\mathcal{E}_F}} \right).
\end{align}

From Eq. (\ref{eq: Mott formula}), we know that the sign of the Seebeck coefficient depends on the sign of the $\frac{d\sigma}{d\mathcal{E}}|_{\mathcal{E}=\mathcal{E}_F}$. So we calculate
\begin{align}
    \frac{d\sigma_{xx}}{d\mathcal{E}_F} = \frac{4e^2\tau}{(2\pi)^2\hbar^2} \sqrt{\frac{m_y}{m_x}} \left( \log \frac{\Lambda_x + \sqrt{\Lambda_x^2 +2\mathcal{E}_F}}{\sqrt{2\mathcal{E}_F}} - \frac{\Lambda_x}{\sqrt{\Lambda_x^2-2\mathcal{E}_F}} \right).
\end{align}

We do the same calculation for the $y$ direction.
\begin{align}
    \sigma_{yy} &= \frac{4e^2\tau\sqrt{m_xm_y}}{(2\pi)^2\hbar^2} \int_0^{\Lambda_x} \int_0^{\sqrt{\Lambda_x^2 - 2\mathcal{E}_F}} d\Tilde{k}_x d\Tilde{k}_y \frac{\Tilde{k}_y^2}{m_y}    \delta \left( \frac{\Tilde{k}_x^2}{2} - \frac{\Tilde{k}_y^2}{2} - \mathcal{E}_F \right) \\
     &= \frac{4e^2\tau\sqrt{m_xm_y}}{(2\pi)^2\hbar^2} \int_{\sqrt{\mathcal{E}_F}}^{\Lambda_x} d\Tilde{k}_x \frac{\sqrt{\Tilde{k}_x^2 - 2\mathcal{E}_F}}{m_y} \\
     &= \frac{4e^2\tau\sqrt{m_xm_y}}{(2\pi)^2\hbar^2} \frac{1}{2}  \left( \Lambda_x \sqrt{\Lambda_x^2 -2\mathcal{E}_F} - 2\mathcal{E}_F \log \frac{\sqrt{\Lambda_x^2 -2\mathcal{E}_F}+\Lambda_x}{\sqrt{2\mathcal{E}_F}} \right).
\end{align}

\begin{equation}
    \frac{d\sigma_{yy}}{d\mathcal{E}_F} = \frac{4e^2\tau}{(2\pi)^2\hbar^2} \sqrt{\frac{m_x}{m_y}} \left( -\log \frac{\Lambda_x + \sqrt{\Lambda_x^2 -2\mathcal{E}_F}}{\sqrt{2\mathcal{E}_F}} \right).
\end{equation}

The condition for ADCP requires $S_{xx}$ and $S_{yy}$ , and therefore $\frac{d\sigma}{d\mathcal{E}}|_{\mathcal{E}=\mathcal{E}_F}$ to have different signs in different directions. From Eq.(\ref{eq: dsigmayy}), we can see that $\frac{d\sigma_{yy}}{d\mathcal{E}} < 0 \implies S_{yy} >0$. Then, this Fermi surface shows ADCP under the assumption $\Lambda_x^2 > 2.31 \mathcal{E}_F$, which ensures $\frac{d\sigma_{xx}}{d\mathcal{E}} > 0 \implies S_{xx} < 0$. 

Therefore, at low temperature and under the assumption mentioned above $k_B T \ll \mathcal{E}_F < \Lambda_x^2/2.31$, a Fermi surface near a saddle point shows ADCP.

\section{Calculation of Seebeck coefficients for intrinsic semiconductors } \label{conductivity-semiconductor}

To calculate the energy dependent conductivity expression for intrinsic semiconductor with parabolic dispersion, we start with a Boltzmann equation and find, analogous to the calculation of the previous section

\begin{equation}
    \sigma_{ii} = \frac{e^2 \tau_e}{(2\pi)^3} \int v_{i}^2 \delta(\mathcal{E} - \mathcal{E}_{\mathbf{k}}) d^3k
\end{equation}

Specifically, consider a parabolic electron dispersion, with different effective masses in different directions:
\begin{equation}
    \mathcal{E} = \frac{1}{2} \left( \frac{\hbar^2k_x^2}{m_{ex}} + \frac{\hbar^2k_y^2}{m_{ey}} + \frac{\hbar^2k_z^2}{m_{ez}} \right).
\end{equation}

Changing variables to $\Tilde{k}_i = \hbar k_i /\sqrt{2m_i}$, our expression becomes 

\begin{equation}
    \sigma_{e,ii} = \frac{e^2 \tau_e 2^{3/2}}{(2\pi\hbar)^3} (m_{ex}m_{ey}m_{ez})^{1/2} \int \frac{2}{m_{ei}} \Tilde{k}_i^2 \delta(\mathcal{E} - \sum_{i=1}^3 \Tilde{k}_i^2) d^3k
\end{equation}

Carrying out the integration, we find

\begin{equation}
\label{eq: energy dependent sigmaexx}
    \sigma^e_{xx} (\mathcal{E}) = \frac{2\sqrt{2}}{3} \frac{e^2}{\pi^2\hbar^3} \left( \frac{m_{ey}m_{ez}}{m_{ex}} \right)^{1/2} \tau_e \mathcal{E}^{3/2}.
\end{equation}
And similarly,
\begin{equation}
\label{eq: energy dependent sigmahxx}
    \sigma^h_{xx} (\mathcal{E}) = \frac{2\sqrt{2}}{3} \frac{e^2}{\pi^2\hbar^3} \left( \frac{m_{hy}m_{hz}}{m_{hx}} \right)^{1/2} \tau_h \mathcal{E}^{3/2}.
\end{equation}
Defining a quasiparticle energy variable in reference to the conduction band bottom $\mathcal{E}_c$, $\mathcal{E}' = \mathcal{E} - \mathcal{E}_c$, in the non-degenerate limit, the Fermi-Dirac distribution can be approximated by the Boltzmann distribution:

\begin{equation}
    f(\mathcal{E}) \propto e^{-\beta (\mathcal{E} - \mu)} \propto e^{-\beta (\mathcal{E}' + \mathcal{E}_c - \mu)} \propto e^{-\beta \mathcal{E}'}e^{-\beta(\mathcal{E}_c - \mu)}.
\end{equation}
\begin{equation}
    -\frac{\partial f}{\partial \mathcal{E}} = \beta e^{-\beta \mathcal{E}'}e^{-\beta (\mathcal{E}_c - \mu)}.
\end{equation}
\begin{align}
\sigma^e_{xx} = \int_0^{\infty} d\mathcal{E}' A_{ex} \mathcal{E}'^{3/2} \tau_e \beta e^{-\beta \mathcal{E}'}e^{-\beta (\mathcal{E}_c - \mu)}, 
\end{align}
where $A_{ex} =  \frac{2\sqrt{2}}{3} \frac{e^2}{\pi^2\hbar^3} \left( \frac{m_{ey}m_{ez}}{m_{ex}} \right)^{1/2} $.
The integral gives us 
\begin{equation}
    \sigma^e_{xx} = A_{ex} \tau_e \beta^{-3/2} e^{-\beta (E_c - \mu)} \Gamma\left(\frac{5}{2}\right).
\end{equation}
Similarly, the Peltier conductivity for the electron pocket gives us 
\begin{align}
    \alpha^e_{xx} &= - \frac{1}{eT} \int_0^{\infty} d\mathcal{E}' A_{ex} \mathcal{E}'^{3/2} \tau_e (\mathcal{E}' + \mathcal{E}_c -\mu)\beta e^{-\beta \mathcal{E}'}e^{-\beta (\mathcal{E}_c - \mu)} \\
    & =  - \frac{1}{eT} A_{ex} \tau_e \beta^{-3/2}e^{-\beta(\mathcal{E}_c - \mu)}\Gamma\left(\frac{5}{2} \right) \left(\frac{5}{2} k_BT + \mathcal{E}_c -\mu \right).
\end{align}
Through similar procedure, we can evaluate the conductivities of the hole pocket using the relative quasiparticle energy from the valance band edge $\mathcal{E}' = \mathcal{E} - \mathcal{E}_v$. The distribution function and the derivative are:
\begin{align}
    f(\mathcal{E}') &= 1 - e^{\beta(\mathcal{E}' + \mathcal{E}_v - \mu)} \\
    -\frac{\partial f}{\partial \mathcal{E}'} &= \beta e^{\beta \mathcal{E}'}e^{\beta(\mathcal{E}_v - \mu)}
\end{align}
The conductivities are in the following:
\begin{align}
    \sigma^h_{xx} &= \int_{-\infty}^0 d\mathcal{E}' A_{hx} (-\mathcal{E}')^{3/2} \tau_h \beta e^{\beta \mathcal{E}'}e^{\beta (\mathcal{E}_v - \mu)} \\
    &= A_{hx} \tau_h \beta^{-3/2} e^{\beta (\mathcal{E}_v - \mu)} \Gamma\left(\frac{5}{2}\right),
\end{align}
where $A_{hx} =  \frac{2\sqrt{2}}{3} \frac{e^2}{\pi^2\hbar^3} \left( \frac{m_{hy}m_{hz}}{m_{hx}} \right)^{1/2} $.
\begin{align}
    \alpha^h_{xx} &= \frac{1}{eT} \int_{-\infty}^0 d\mathcal{E}' A_{hx} (-\mathcal{E}')^{3/2} \tau_h (\mathcal{E}' + \mathcal{E}_v -\mu)\beta e^{\beta \mathcal{E}'}e^{\beta (\mathcal{E}_v - \mu)} \\
    &= \frac{1}{eT} A_{hx}\tau_h \beta^{-3/2}e^{\beta (\mathcal{E}_v - \mu)}\Gamma\left(\frac{5}{2}\right) \left( \frac{5}{2} k_BT +\mu - \mathcal{E}_v \right).
\end{align}

Using Eqs. (\ref{eq: je}-\ref{generalS}) and with the assumption that the gap $\Delta$ is small, and the chemical potential is in the middle of the gap,i.e., $(\mathcal{E}_c - \mu = \mu - \mathcal{E}_v = \Delta/2)$, we can find, 
\begin{equation}
    S_{xx} = -\frac{1}{eT} \frac{A_{ex}\tau_e- A_{hx}\tau_h}{A_{ex}\tau_e+ A_{hx}\tau_h} \left( \frac{5}{2} k_BT + \frac{\Delta}{2} \right).
\end{equation}
Similarly,
\begin{equation}
    S_{yy} = -\frac{1}{eT} \frac{A_{ey}\tau_e- A_{hy}\tau_h}{A_{ey}\tau_e+ A_{hy}\tau_h} \left( \frac{5}{2} k_BT + \frac{\Delta}{2} \right).
\end{equation}

\section{Condition for ADCP for Semiconductor with Energy-dependent Relaxation Times}\label{appendix: energy dependent relaxation times}

When we derive the condition for ADCP for intrinsic semiconductors with energy dependent relaxation times, we find the expressions to be more complex and include the band gap and temperature. 

We write the relaxation times as 
\begin{align}
\tau_e &= \tau_{e0} |\mathcal{E}_F-\mathcal{E}_{e,0}|^{p_e -1} \\
\tau_h &= \tau_{h0} |\mathcal{E}_F-\mathcal{E}_{h,0}|^{p_h -1},
\end{align}
 and then we follow the same calculation as Appendix \ref{conductivity-semiconductor}. We derive that the condition for the material to exhibit ADCP, i.e., for it to have $S_{ii}>0$ and $S_{jj}<0$ in $i$ and $j$ directions is:
 \begin{multline}
  \left( \frac{A_{ei}\tau_{e0}}{A_{hi}\tau_{h0}} (k_B T)^{p_e -p_h} \frac{\Gamma\left(p_e+\frac{3}{2}\right)}{\Gamma\left(p_h+\frac{3}{2}\right)} - \frac{\left(p_h+\frac{3}{2}\right)k_BT + \frac{\Delta}{2} }{\left(p_e+\frac{3}{2}\right)k_BT + \frac{\Delta}{2}}\right) \left( \frac{A_{ej}\tau_{e0}}{A_{hj}\tau_{h0}} (k_B T)^{p_e -p_h} \frac{\Gamma\left(p_e+\frac{3}{2}\right)}{\Gamma\left(p_h+\frac{3}{2}\right)} - \frac{\left(p_h+\frac{3}{2}\right)k_BT + \frac{\Delta}{2} }{\left(p_e+\frac{3}{2}\right)k_BT + \frac{\Delta}{2}}\right) < 0.
\end{multline}
Here, $A_{e/h,i} =  \frac{2\sqrt{2}}{3} \frac{e^2}{\pi^2\hbar^3} \frac{\left( m_{e/h} \right)^{1/2}}{m_{e/h,i}} $ and $m_{e/h} = \Pi_{i=x,y,z} m_{e/h,i}$.
    
\end{widetext}

\bibliography{refs}

\end{document}